\documentclass[twocolumn,superscriptaddress,amsmath,amssymb,aps,pra]{revtex4}
\usepackage{amsmath,amsthm,amssymb}
\usepackage{latexsym}
\usepackage{amscd,graphicx}
\usepackage[titletoc]{appendix}
\usepackage{epsfig}
\usepackage{epstopdf}
\usepackage[normalem]{ulem}
\usepackage[dvipsnames]{xcolor}
\usepackage[colorlinks=true,linkcolor=red,citecolor=blue]{hyperref}

\usepackage{pdfpages}
\usepackage{subfigure}
\usepackage{tabularx}

\newcommand{\abs}[1]{| #1 |}

\begin{document}
	\title{Nonreciprocal Entanglement in Cavity-Magnon Optomechanics}
	
	\author{Jiaojiao Chen}
	\affiliation{Department of Physics and Optoelectronic Engineering,Anhui University, Anhui 230000, China}
	
	\author{Xiao-Gang Fan}
	\affiliation{Department of Physics and Optoelectronic Engineering,Anhui University, Anhui 230000, China}
	
	\author{Wei Xiong}
	%\altaffiliation{These authors equally contribute to this work}
	\altaffiliation{xiongweiphys@wzu.edu.cn}
	\affiliation{Department of Physics, Wenzhou University, Zhejiang 325035, China}
	
	\author{Dong Wang}
	\affiliation{Department of Physics and Optoelectronic Engineering,Anhui University, Anhui 230000, China}
	
	\author{Liu Ye}
	\altaffiliation{yeliu@ahu.edu.cn}
	\affiliation{Department of Physics and Optoelectronic Engineering,Anhui University, Anhui 230000, China}

	\date{\today }
	
	\begin{abstract}
		Cavity optomechanics, a promising platform to investigate macroscopic quantum effects, has been widely used  to study nonreciprocal entanglement with Sagnac effect. Here we propose an alternative way to realize nonreciprocal entanglement among magnons, photons, and phonons in a hybrid cavity-magnon optomechanics, where magnon Kerr effect is used. We show that the Kerr effect gives rise to a magnon frequency shift and an additional two-magnon effect. Both of them can be tuned from positive to negative via tuning the magnetic field direction, leading to nonreciprocity. By tuning system parameters such as magnon frequency detuning or the coefficient of the two-magnon effect, bipartite and tripartite entanglements can be nonreciprocally enhanced. By further studying the defined bidirectional contrast ratio, we find that nonreciprocity in our system can be switched on and off, and can be engineered by the bath temperature. Our proposal not only provides a potential path to demonstrate nonreciprocal entanglement with the magnon Kerr effect, but also opens a direction to engineer and design diverse nonreciprocal devices in hybrid cavity-magnon optomechanics with nonlinear effects.	
	\end{abstract}
	
	\maketitle
	\section{Introduction}
	Macroscopic quantum entanglement, as a core resource in quantum information science~\cite{Bouwmeester-2000}, is crucial to understand the classical-to-quantum boundary~\cite{Haroche-1998}. Such entanglement is generally generated in bilinear or nonlinear quantum systems. Cavity optomechanics (COM)~\cite{Aspelmeyer}, formed by photons nonlinearly coupled to phonons via radiation pressure, is a promising platform to investigate quantum effects~\cite{Xiong-2021,Lu-2013,Chen-2021,Xiong-2023,Xiong3-2016,Lu-2015,Xiong3-2021,Xiong4-202205}, especially for macroscopic quantum effects theoretically~\cite{Vitali-2007,Tian-2013,Wang-2013,Mancini-2002,Lai-2022} and experimentally~\cite{Palomaki-2013,Korppi-2018,Kotler-2021,de Lepinay-2018,Riedinger-2018}. Very recently, nonreciprocal entanglement in COM has attracted great interest~\cite{Jiao-2020,Jiao-2022}. This is because entanglement can be well protected (enhanced) by breaking the Lorentz reciprocity~\cite{Jiao-2020}. Utilizing this, various nonreciprocal devices in COM have been realized~\cite{Xu-2019,Shen-2016,Lu-2021,Yao-2022,Jiang-2018,Peng-2023,Xie-2022,Li-2020}. Previous proposals for studying nonreciprocal entanglement in COM~\cite{Jiao-2020,Jiao-2022} mainly rely on Sagnac effect~\cite{Malykin-2000,Maayani-2018}, which causes a positive or negative shift on the cavity resonance frequency, dependent on the direction of the driving field on the cavity. Apart from COM, magnons in the Kittle mode of ferromagnetic yttrium-iron-garnet (YIG) spheres~\cite{Rameshti-2022,Yuan-2022,Quirion-2019,Wang-2020,Zheng-2023} can also provide new insights for studying macroscopic quantum effects~\cite{Li-2018,Yu-2020,Zhang-2019}. This is due to the fact that magnons have an intrinsic Kerr effect from the magnetocrystallographic anisotropy~\cite{Shen-2021,Shen-2022}, which can also give a positive or negative frequency shift on the Kittle mode by tuning the direction of magnetic field~\cite{Wang-2016,Wang-2018,ZhangGQ-2019}. The Kerr effect has been employed to investigate various phenomena, including multistability~\cite{Shen-2021,Shen-2022,Wang-2018}, long-distance spin-spin interaction~\cite{Xiong1-2022}, quantum phase transition~\cite{ZhangGQ-2021,Liu-2023}, and sensitive detection~\cite{ZhangGQ-2023}. However, nonreciprocal entanglement has not yet been revealed with the Kerr effect.
	
	{Here} we propose how to realize nonreciprocal bi- and tripartite entanglements in a hybrid cavity-magnon optomechanics. We find that, not only all bipartite entanglements but also a genuine tripartite entanglement can be generated in the absence of magnon Kerr effect, and the initial optomechanical entanglement can be partially transferred to the cavity-magnon and magnon-phonon subsytems.  When the Kerr effect is considered,  a mean magnon-number-dependent frequency shift on magnons is produced. Similar to Sagnac effect~\cite{Malykin-2000,Maayani-2018} on the cavity field, the Kerr effect induced frequency shift can be positive or negative by tuning the direction of the magnetic field. Different from Sagnac effect, the magnon Kerr effect also gives rise to an additional two-magnon effect, which modulates the maximum values of all entanglements in our setup. As a result, both the optomechanical and magnon-phonon entanglements are reduced, but magnon-photon and the tripartite entanglements are enhanced, compared to the case without Kerr effect. By further tuning the aligned magetic field along the crystallographic axis $[100]$ or $[110]$, one can see that all entanglements can be nonreciprocally generated. Interestly, all entanglements except for the optomechanical entanglement can be nonreciprocally enhanced with accessible parameters. This indicates that entanglement transfer from the optomechancial entanglement to the cavity-magnon and magnon-phonon subsytems is nonreciprocal. Finally, we show that perfect nonreciprocity for all bi-and tripartite entanglements can be achieved, by studying the defined bidirectional contrast ratio. {The achieved bi-and tripartite entanglements in our proposal are continuous variable (CV) entanglements, which has been widely applied to quantum transduction~\cite{Tian-2022,Zhong-2022}, quantum networking ~\cite{Cirac-1997,Lodahl-2017,Gonzalez-Ballestero-2015,Gangaraj-2017,Hu-2019,Kimble-2008}, quantum sensing ~\cite{Degen-2017}, Bell-state test~\cite{Marinkovic-2018}, quantum teleportation ~\cite{Hofer-2011,Hofer-2015,Barzanjeh-2012}, microwave-optics conversion ~\cite{Xu-2016,Tian-2017,Eshaqi-Sani-2022,Ren-2022}, and other CV quantum information processing~\cite{Andersen-2010,Stannigel-2012,Horodecki-2009,Braunstein-2005}. Thus, CV entanglement can be regarded as a useful resource for CV quantum information science.} Our work provides a potential way to nonreciprocally enhance and engineer quantum entanglement with Kerr effect, and opens a promising path to realize diverse nonreciprocal devices with magnon Kerr effect.

	\begin{figure}
		\includegraphics[scale=0.6]{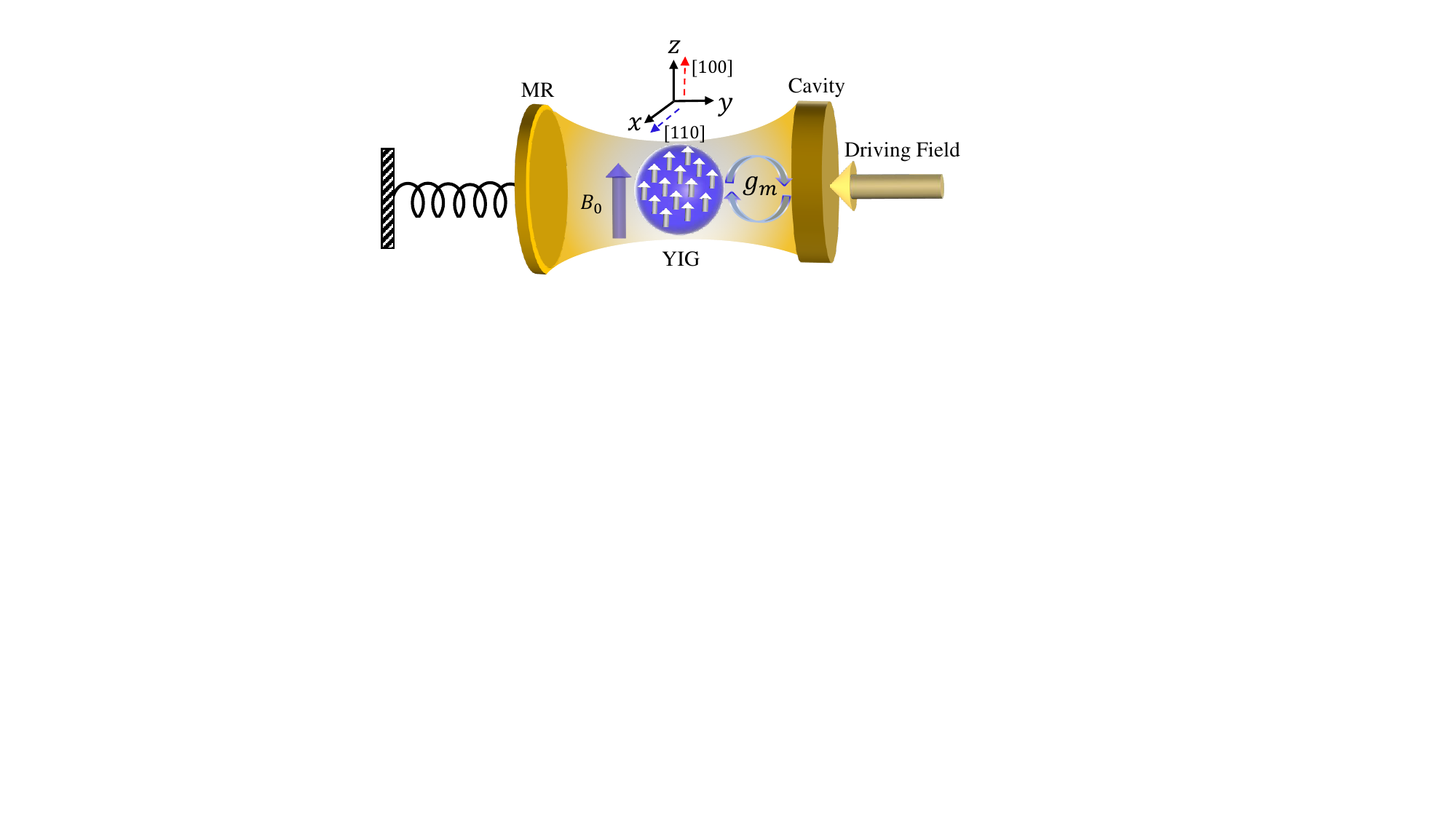}
		\caption{(a) Schematic diagram of the proposed cavity-magnon optomechanical system. It consists of a driven cavity simultaneously coupled to both a Kittle mode with Kerr nonlinearity in a YIG sphere and a MR, with coupling strength $g_m$ and $g_0$, respectively. The YIG sphere is placed in a static magnetic field, along the crystallographic axis $[100]$ (see the red-arrowed line) or $[110]$ (see the blue-arrowed line) of the YIG sphere.}
		\label{fig1}
	\end{figure}
	
	\section{Model and Hamiltonian}
	
	We consider a hybrid cavity-magnon optomechanical system consisting of a strongly driven cavity with frequency $\omega_a$ coupled to both a mechanical resonator (MR) with frequency $\omega_b$ and a micron-size YIG sphere supporting a Kittle mode with frequency $\omega_m$, where the YIG sphere is positioned in a static magnetic field $B_0$~(see Fig.~\ref{fig1}). The Hamiltonian of the proposed hybrid system can be written as (setting $\hbar=1$) $H=H_{\rm om}+H_{\rm kerr}+H_{\rm int}+H_d$, where the COM Hamiltonian, $H_{\rm om}=\omega_a a^\dag a+\dfrac{1}{2}\omega_b(p^2+q^2)-g_{0} a^\dagger aq$, describes the radiation pressure interaction between the cavity field and the MR. $g_0=(\omega_a/L)\sqrt{\hbar/m\omega_b}$ is the single-photon optomechanical coupling strength, with $L$ being the cavity length in the absence of the intracavity field and $m$ the effective mass of the MR. The second term $H_{\rm kerr}=\omega_m m^\dag m+K_0 (m^\dag m)^2$ characterizes the Kerr nonlinearity of magnons in the Kittle mode, where $K_0$ is reversely propotional to the {volume} of the YIG sphere, and its sign can be tuned by varying the direction of the static magnetic field. Specifically, when the crystallographic axis $[100]$ ($[110]$) is aligned along the magnetic field, $K_0>0~(<0)$~\cite{Wang-2018}. Experimentally, $K_0$ can be tuned from $0.05$ nH to $100$ nH for the diameter of the YIG sphere from $1$ mm to $100$ ${\rm \mu}$m. Here, $a$~($a^\dag$) and $m$ ($m^\dagger$) are the anihilation (creation) operators of the cavity  and Kittle modes, respectively, and $q$~($p$) is the dimensionless position~(momentum) quadrature of the MR. The Hamiltonian $H_{\rm int}=g_m (a^\dag m+a m^\dag)$ represents the magnetic dipole coupling between the cavity and the Kittle mode with the tunable coupling strength $g_m$. The last term, $H_d=i\Omega_0(a^\dag e^{-i\omega_0 t}-a e^{i\omega_0 t})$ with frequency $\omega_0$ and Rabi frequency ${\Omega_0}$, denotes the coupling between the driving field and the cavity. With the strong driving field ($\Omega_0\gg\kappa_a,\gamma_m$), the higher-order fluctuation terms in quantum Langevin equations can be safely neglected when each operator is rewritten as the steady-state value plus its quantum fluction, i.e., $O\rightarrow O_s+O$, with $O=a,m,q,p$.  Then the linearized Hamiltonian of the hybrid system is given by (see the detailed derivation in the Appendix) 
	\begin{align}\label{q1}
	\mathcal{H}=&\tilde{\Delta}_a a^\dag a+\tilde{\Delta}_m m^\dag m+\frac{1}{2}\omega_b (q^2+p^2)-\frac{1}{\sqrt{2}}{g_b}(a+a^\dag)q\notag\\
	&+g_m (a^\dag m+a m^\dag)+\frac{1}{2}K(m^{\dag 2}+m^2),
	\end{align}
	where $\tilde{\Delta}_a=\omega_{a}-\omega_0-g_0 q_s$ is the effective cavity frequency detuning induced by the displacement $q_s$ of the MR, and $\tilde{\Delta}_m=\Delta_m+\Delta_K$, with the magnon frequency detuning $\Delta_m=\omega_m-\omega_0$ and the magnon frequency shift $\Delta_K\equiv2K$, is the effective magnon frequency detuning induced by magnon Kerr effect. The defined parameter $K\equiv2K_0 N_m$ characterizes the strength of the two-magnon effect, which can squeeze magnons~\cite{Xiong1-2022}. As $K_0$ can be positive (negative), so $K>0$~($<0$), leading to $\Delta_K>0$~($<0$). Obviously, $K$ can be significantly amplified by the steady-state magnon number $N_m=|m_s|^2$, which can be indirectly tuned by the strong driving field acting on the cavity, via the beam-splitter interaction between the cavity and the Kittle mode (i.e., $a^\dag m+a m^\dag$). $g_b\equiv \sqrt{2}g_0a_s$ is the effective linearized optomechanical coupling strength, directly tuned by the strong driving field acting on the cavity (i.e., $a_s\propto \Omega_0$). For simplicity,  $m_s$ is assumed to be real via properly choosing the phase of the driving field.

	\section{Dynamics and entanglement metric}
	
	According to quantum Langevin equation, the dynamics of the linearized hybrid system governed by the Hamiltonian~(\ref{q1}) can be written as (see details in the Appendix)
	\begin{align}\label{q2}
	\dot{q}=&\omega_b p,~~\dot{p}=-\omega_b q+g_b(a+a^\dag)/\sqrt{2}-\gamma_b p+\xi,\notag\\
	\dot{a}=&-(i\tilde{\Delta}_a+\kappa_a) a+ig_b q/\sqrt{2}-ig_m m+\sqrt{2\kappa_a} a_{\rm in},\\
	\dot{m}=&-(i\tilde{\Delta}_m+\gamma_m) m-ig_m a-iKm^\dag+\sqrt{2\gamma_m} m_{\rm in},\notag
	\end{align}   
	where $a_{\rm in},~m_{\rm in}$, and~$\xi $ are the input noise operators with zero mean value (i.e., $\langle a_{\rm in}\rangle=\langle m_{\rm in}\rangle=\langle \xi\rangle=0$). Under the Markovian approximation, two-time correlation functions of these input noise operators in the resolved sideband regime (i.e., $\omega_b\gg\gamma_b$) are given by~\cite{Walls-1994}	$\langle a_{\rm in}^\dagger(t^\prime)  a_{\rm in}(t)\rangle=\bar{n}_{a}\delta(t-t^\prime),	\langle a_{\rm in}(t)  a_{\rm in}^\dagger(t^\prime)\rangle=(\bar{n}_{a}+1)\delta(t-t^\prime),
	\langle m_{\rm in}^\dagger(t^\prime)  m_{\rm in}(t)\rangle=\bar{n}_{m}\delta(t-t^\prime),
	\langle m_{\rm in}(t)  m_{\rm in}^\dagger(t^\prime)\rangle=(\bar{n}_{m}+1)\delta(t-t^\prime),
	\langle \xi(t)\xi(t^\prime)+\xi(t^\prime)\xi(t)\rangle/2\simeq\gamma_b(2\bar{n}_{b}+1)\delta(t-t^\prime),$
	where $ \bar{n}_\sigma=[{\rm exp}({\hbar\omega_\sigma}/{k_{B}T}-1)]^{-1}~(\sigma=a,b,m)$, with $k_B$ being the Boltzmann
	constant and $T$ the bath temperature, are the mean thermal excitation number in the cavity, the Kittle mode, and the MR, respectively. In a compact form, Eq.~(\ref{q2}) can be rewritten as $\dot{u}(t)=Au(t)+f(t)$, where $u(t)=[x_a (t), y_a (t), x_m (t),y_m (t),  q(t),  p(t)]^T$ and $f(t)=[\sqrt{2\kappa_a} x_{\rm in}^{a}(t), \sqrt{2\kappa_a} y_{\rm in}^{a}(t), \sqrt{2\gamma_m} x_{\rm in}^{m}(t),
	\sqrt{2\gamma_m} y_{\rm in}^{m}(t), 0, \xi(t)]^T$ are the vectors of the system and the input noise operators, respectively, and the drift (coefficient) matrix $A$ is (see details in the Appendix)
	\begin{equation}\label{q3}
	A=\left(\begin{array}{cccccc}
	-\kappa_a  &\tilde{\Delta}_a&0&g_m& 0&  0\\
	-\tilde{\Delta}_a  &-\kappa_a&-g_m &0 &g_b&0\\
	0  &g_m &-\gamma_m&\tilde{\Delta}_m^-&0 &0\\
	-g_m&0&-\tilde{\Delta}_m^+&-\gamma_m&0 &0\\
	0 &0&0 &  0  & 0  &\omega_b\\
	g_b &0&0&0 & -\omega_b & -\gamma_b
	\end{array}\right),
	\end{equation}
	where $\tilde{\Delta}_m^\pm=\tilde{\Delta}_m\pm \Delta_K/2$.
	\begin{figure}
		\includegraphics[scale=0.35]{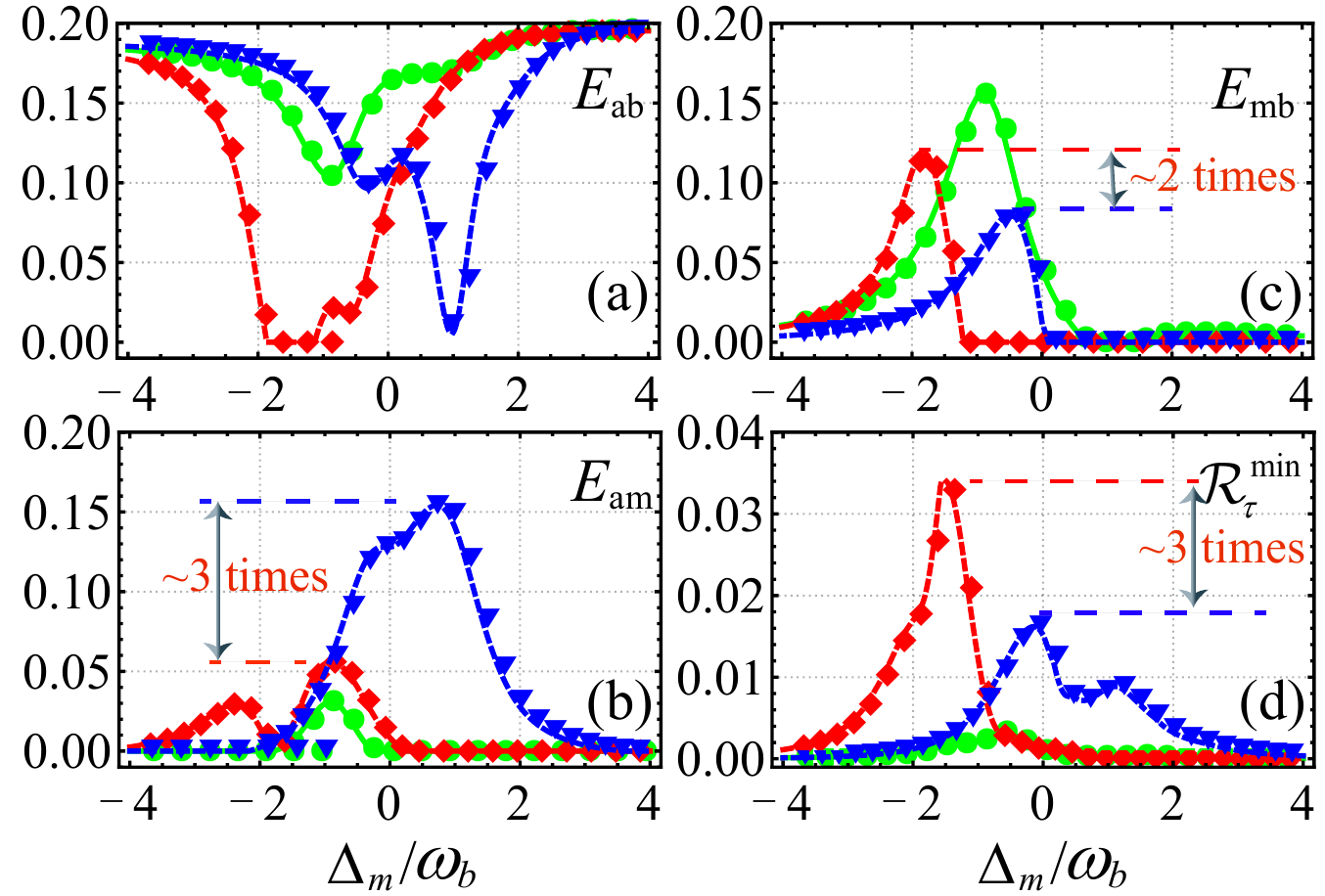}
		\caption{Logarithmic negativities (a) $E_{\rm ab}$, (b) $E_{\rm am}$, (c) $E_{\rm mb}$,  and (d) the minimum residual contangle $\mathcal{R}_{\tau}^{\rm min}$, versus magnon frequency detuning $\Delta_m$ with $\Delta_K>0$ (the red curve), $\Delta_K=0$ (the green curve), and $\Delta_K<0$ (the blue curve). The parameters are $\omega_a=\omega_m=2\pi\times10$~GHz, $\omega_b/2\pi=10$~MHz, $\kappa_a/2\pi=\gamma_m/2\pi=0.4\omega_b$, $\gamma_b/2\pi=100$~Hz, $g_m=g_b=2\pi\times0.5\omega_b$, $K=\kappa_a$, $T=10$~mK, and $\tilde{\Delta}_a=\omega_b$.}
		\label{fig2}
	\end{figure}
	
	Since the input quantum noises are zero-mean quantum Gaussian noises, the quantum steady state for the fluctuations is a zero-mean CV three-mode Gaussian state, fully characterized by a $ 6\times6 $ covariance matrix $
	\mathcal{V}_{ij}=\langle u_{i}(t)u_{j}(t^\prime)+u_{j}(t^\prime)u_{i}(t)\rangle/2$ $(i,j=1,2,...,6)$, where the steady-state $\mathcal{V}$ can be given by solving the Lyapunov equation
	\begin{align}\label{eq11}
	A\mathcal{V}+\mathcal{V}A^T=-D.
	\end{align}
	Here $ D={\rm diag} [\kappa_a(2\bar{n}_a+1)$, $\kappa_a(2\bar{n}_a+1), \gamma_m(2\bar{n}_m+1)$, $\gamma_m(2\bar{n}_m+1), 0, \gamma_b(2\bar{n}_b+1)]$  is defined by $ \langle n_{i}(t)n_j(t^\prime)+n_j(t^\prime)n_{i}(t)\rangle/2=D_{ij}\delta(t-t^\prime)$. To investigate bipartite and tripartite entanglement of the proposed system, the logarithmic negativity $E_N$~\cite{Vidal-2002,Plenio-2005} and the residual contangle $R_\tau$~\cite{Adesso-2006} are employed, respectively. A {\it bona fide} quantification of tripartite entanglement is given by the {\it minimum} residual contangle~\cite{Adesso-2006}, $\mathcal{R}_\tau^{\rm min}\equiv{\rm min}[\mathcal{R}_\tau^{m|ab},\mathcal{R}_\tau^{b|am},\mathcal{R}_\tau^{a|mb}]$, where $ R_\tau^{i|jk}\equiv C_{i|jk}-C_{i|j}-C_{i|k}\geq0 $ $(i,j,k=a,m,b)$, with $ C_{u|v
	} $ being the contangle of subsystem of $ u $ and $ v $ ($ v $ contains one or two modes), is a proper entanglement monotone defined as the sqaured logarithmic negativity. A nonzero minimum residual contangle $\mathcal{R}_\tau^{\rm min}>0$ means the presence of genuine tripartite entanglement in the system.

	\section{Nonreciprocal entanglement with Kerr effect}
	
	Before starting, we first point out that nonreciprocal entanglement induced by the magnon Kerr effect is different from the mechanism of the Sagnac effect. This is due to the fact that, the magnetic field mediated Kerr effect not only gives a red (blue) shift in magnon frequency, but also generates a two-magnon effect. To study nonreciprocal entanglement,  the experimentally accessible parameters are used: $\omega_a=\omega_m=2\pi\times10$~GHz, $\omega_b/2\pi=10$~MHz, $\kappa_a/2\pi=\gamma_m/2\pi=0.4\omega_b$, $\gamma_b/2\pi=100$~Hz, $g_m=g_b=2\pi\times0.5\omega_b$, $K=\kappa_a$, $T=10$~mK, $\tilde{\Delta}_a=\omega_b$, and $\Delta_m=-\omega_b$. These parameters numerically guarantee the system stable according to the Routh-Hurwitz
	criterion~\cite{RH}. To investigate nonreciprocal entanglements, we plot three logarithmic negativities and the minimum residual contangle versus the magnon frequency detuning $\Delta_m$ in Fig.~\ref{fig2}. The red and blue curves respectively denote the magnetic field along the crystalline axis $[100]$ and $[110]$, corresponding to $\Delta_K>0$ and $\Delta_K<0$. For comparison, entanglement without the Kerr effect (i.e., $\Delta_K=0$) is also presented [see the green curve in Fig.~\ref{fig2}]. From Fig.~\ref{fig2}(a), we can see that the optomechanical entanglement $E_{\rm ab}$  decreases first and then increases with $\Delta_m$ in the absence of the Kerr effect [see the green curve], while  magnon-photon ($E_{\rm am}$) and magnon-phonon ($E_{\rm mb}$) entanglements increase first and then decrease [see Figs.~\ref{fig2}(b) and \ref{fig2}(c)], which is fully opposite to $E_{\rm ab}$. This indicates that the initial magnon-phonon entanglement is partially transferred to the cavity-magnon and -phonon subsystems, owing to the mediation of photons. Besides, a {\it genuinely} tripartite entanglement is generated around $\Delta_m\approx-\omega_b$, as demonstrated by the nonzero minimum residual contangle $\mathcal{R}_{\tau}^{\rm min}$ in Fig.~\ref{fig2}(d). When the Kerr effect is taken into account, both $E_{\rm ab}$ and $E_{\rm mb}$ have a certain reduction, but $E_{\rm am}$ and $\mathcal{R}_{\tau}^{\rm min}$ are enhanced. By tuning the direction of the magnetic field, i.e., changing $\Delta_K>0~(<0)$ to $\Delta_K<0~(>0)$,  all entanglements have different responses~[see red and blue curves in Fig.~\ref{fig2}], corresponding the nonreciprocity. Utilizing this nonreciprocity, magnon-phonon, magnon-photon and magnon-photon-phonon entanglements can be enhanced by $\sim 2$, $3$ and $3$ times, respectively.
	\begin{figure}[h]
		\centering
		\includegraphics[scale=0.3]{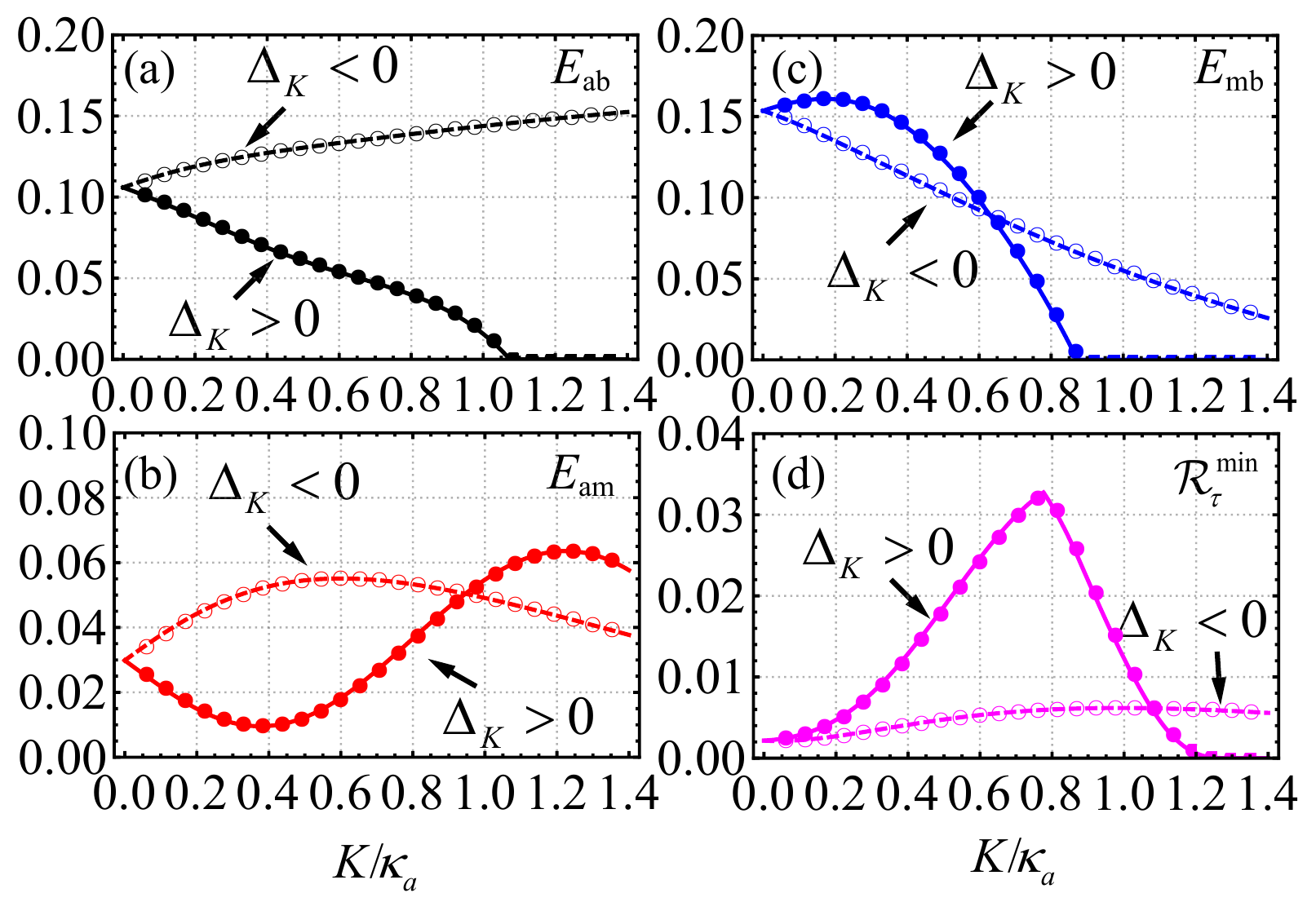} 
		\caption{Logarithmic negativities (a) $E_{\rm ab}$, (b) $E_{\rm am}$, (c) $E_{\rm mb}$, and (d) the minimum residual contangle $\mathcal{R}_{\tau}^{\rm min}$, versus the effective Kerr strength $K$ with $\Delta_K>0$ and $\Delta_K<0$. The solid circles in panels (a)-(d) denote $\Delta_K>0$, and the empty circles denote $\Delta_K<0$. Other parameters are the same as in Fig.~\ref{fig2} except for $\Delta_m/\omega_b=-1$.}
		\label{fig3}
	\end{figure}
	We also plot all entanglements versus the effective Kerr strength $K$ in Fig.~\ref{fig3} to investigate effects of  the Kerr nonlinearity and the magnetic field direction on entanglements. The parameters are the same as those in Fig.~\ref{fig2} but $\Delta_m=-\omega_b$. From Fig.~\ref{fig3}(a), we can see that $E_{\rm ab}$ nearly has a linear dependece on the strength of the Kerr effect for both $\Delta_K>0$ and $\Delta_K<0$. But it monotonously decreases (increases) when $\Delta_K>0~(<0)$.  Figure \ref{fig3}(b) shows that $E_{\rm am}$ is nonlinearly dependent on $K$. Specifically, $E_{\rm am}$ first decreases (increases) and then increases (decreases) when $\Delta_K>0~(<0)$. For $E_{\rm mb}$ in Fig.~\ref{fig3}(c), we find it is linear dependent on $K$ when $\Delta_K<0$, but when $\Delta_K>0$, the dependece becomes nonlinear [see blue solid curve], that is,  $E_{\rm mb}$ decreases slowly when $\Delta_K>0$ than the case of $\Delta_K<0$ first, then the situation becomes opposite passing through the crosspoint. For $\mathcal{R}_{\tau}^{\rm min}$ in Fig.~\ref{fig3}(d), we can see that it is nearly unchanged with $K$ for $\Delta_K<0$, but sharply increases to the maximal value and then decreases for $\Delta_K<0$. These results indicates that all entanglements can be nonreciprocally enhanced with the Kerr effect.
	
	\section{Switchable nonreciprocity}
	\begin{figure}[h]
		\centering
		\includegraphics[scale=0.34]{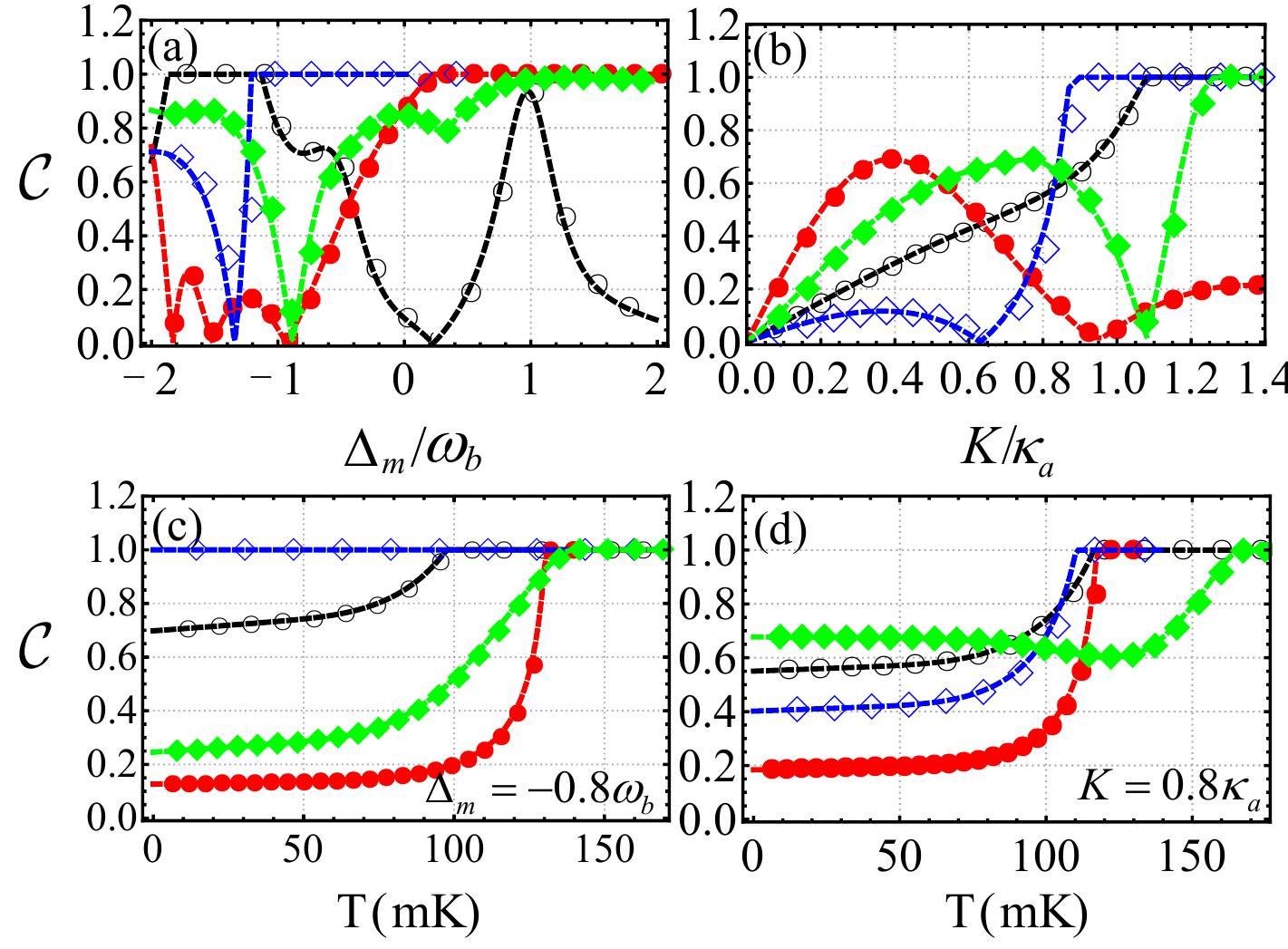} 
		\caption{(a) Bidirectional contrast ratio $\mathcal{C}$ for three bipartite and tripartite entanglements as functions of (a) magnon frequency detuning $\Delta_m$ and (b) the effective strength $K$. The parameters are the same as those in Fig.~\ref{fig2}.}
		\label{fig4}
	\end{figure}
	
	In order to quantitatively describe nonreciprocal entanglement, we introduce the bidirectional contrast ratio $\mathcal{C}$ (satisfying $0\leq\mathcal{C}\leq 1$) for bipartite and tripartite entanglements in the nonreciprocal regimes,
	\begin{align}\label{q6}
	\mathcal{C}_E^{ij}=&\frac{\abs{E_{ij}(>0)-E_{ij}(<0)}}{E_{ij}(>0)+E_{ij}(<0)},\notag\\
	\mathcal{C}_\mathcal{R}=&\frac{|\mathcal{R}_{\tau}^{\rm min}(>0)-\mathcal{R}_{\tau}^{\rm min}(<0)|}{\mathcal{R}_{\tau}^{\rm min}(>0)+\mathcal{R}_{\tau}^{\rm min}(<0)},
	\end{align}
	where $\mathcal{C}_{E}^{ij} ~(\mathcal{C}_{\mathcal{R}})=1$ and $0$ corresponds to the ideal and no nonreciprocities for bipartite (tripartite) entanglements. The higher the contrast ratio $\mathcal{C}$ is, the stronger nonreciprocity of entanglement is. To clearly show this, we numerically plot the contrast ratio $\mathcal{C}$ versus the frequency detuning ($\Delta_m$)  in Fig.~\ref{fig4}(a), where the black, red, blue, and green curves respectively denote the bidirectional contrast ratios $\mathcal{C}_E^{ab}$, $\mathcal{C}_E^{am}$, $\mathcal{C}_E^{mb}$,  and $\mathcal{C}_\mathcal{R}$. Obviously, the nonreciprocity of all bipartite and tripartite entanglements can be switched off and on by tuning  $\Delta_m$ for $K=\kappa_a$. Moreover, the bidirectional contrast ratios for all entanglements can be tuned from $0$ to $1$ by varying $\Delta_m$. This indicates that all entanglements with ideal nonreciprocity can be achieved in our proposal, via tuning the magnon frequency detuning. In Fig.~\ref{fig4}(b), we further study the effect of Kerr strength $K$ on the bidirectional contrast ratios at $\Delta_m=-\omega_b$. It is clearly shown that all entanglements are reciprocal in the absence of the Kerr effect, i.e., $K=0$. When the Kerr effect is considered, all entanglements become nonreciprocay, even for the weak Kerr effect (e.g., $K=0.2\kappa_a$). For the strong Kerr effect (e.g., $K=1.2\kappa_a$), the bidirectional contrast ratios $\mathcal{C}_E^{\rm ab}=\mathcal{C}_E^{\rm am}=\mathcal{C}_\mathcal{R}=1$ can be obtained, while $\mathcal{C}_E^{\rm ab}<1$ in the whole region. This shows that the nonreciprocities of the bipartite entanglements including magnon-photon and magnon-phonon entanglements and the genuinely tripartite entanglements can have ideal nonreciprocities via tuning the strength of the effective Kerr effect. Similar to the case of tuning $\Delta_m$, the nonreciprocities for all entanglements can also be switched off and on with the Kerr effect. In addition, we exame the effect of temperature on the bidirectional contrast ratios with different parameters in  Figs.~\ref{fig4}(c) and \ref{fig4}(d). We find the nonreciprocity of the magnon-phonon entanglement is robust against the temperature when $\Delta_m=-0.8\omega_b$ and $K=\kappa_a$~[see blue curve in Fig.~\ref{fig4}(c)], while nonreciprocities of other entanglements increases slowly first with $T$. By further increasing $T$, a sharp increase occurs and the ideal nonreciprocal photon-phonon, magnon-photon, and magnon-photon-phonon can be achieved~[see other curves in Fig.~\ref{fig4}(c)]. When  $\Delta_m=-\omega_b$ and $K=0.8\kappa_a$~[see curves in Fig.~\ref{fig4}(d)], one can see that nonreciprocities of all entanglements has similar behaviors with the case of the magnon-photon entanglement in Fig.~\ref{fig4}(c). The findings suggest that higher temperature is benifical to obtain the large or optimal nonreciprocity for entanglement, providing another promising path to engineer the nonreciprocity.
	
	\section{Conclusion}
	
	We have proposed a scheme to realize nonreciprocal entanglements with magnon Kerr effect among magnons, photons, and phonons in a hybrid cavity-magnon optomechanical system. By applying a strong driving field on the cavity, Kerr effect gives rise to a positive (negative) frequency shift in the magnon frequency and an additional two-magnon effect. The signs of the frequency shift and the coefficient of the two-magnon effect are dependent on the direction of the applied magnetic field, leading to nonreciprocal entanglements. By further tuning the system parameters, such as the magnon frequency detuning and strength of the Kerr effect, we find entanglements among magnons, photons, and phonons can be nonreciprocally enhanced via changing the direction of the magnetic field. We also show that entanlement nonreciprocity in our proposal, characterized by the defined bidirectional contrast ratio, can be swiched off and on by tuning system parameters. With proper parameters, ideal nonreciprocal entanglements can be achieved. Finally, we find that nonreciprocity can be improved with the bath temperature, even to the ideal value. The results suggest that our scheme provides an alternative path to realize nonreciprocal entanglement with Kerr effect and engineer nonreciprocity with bath temperature. 
	
	This work is supported by the National Natural Science Foundation of China (Grants No.~12175001 and No.~12075001) and the key program of the Natural Science Foundation of Anhui (Grant No. KJ2021A1301).
	
	\setcounter{equation}{0}
	\renewcommand{\theequation}{A\arabic{equation}}
	\appendix{}
	
	\section*{APPENDIX}\label{app}
	
	In this Appendix, we give a detailed derivation of the effective Hamiltonian (\ref{q1}) and the drift matrix $A$ given by Eq.~(\ref{q3}) in the main text. 
	
	We consider a hybrid cavity-magnon optomechanical system consisting of a driven cavity simultanesouly coupled to a micron-sized yttrium-iron-garnet (YIG) sphere and a mechanical resonator. The Hamiltonian of the total system can be written as (setting $\hbar=1$) 
	\begin{equation}
	H=H_{\rm om}+H_{\rm kerr}+H_{\rm int}+H_d\label{eq1}
	\end{equation}
	with
	\begin{align}
	H_{\rm om}=&\omega_a a^\dag a+\dfrac{1}{2}\omega_b(p^2+q^2)-g_{0} a^\dagger aq,\notag \\
	H_{\rm kerr}=&\omega_m m^\dag m+K_0 (m^\dag m)^2 \\
	H_{\rm int}=&g_m (a^\dag m+a m^\dag)\notag \\
	H_d=&i\Omega_0(a^\dag e^{-i\omega_0 t}-a e^{i\omega_0 t})\notag.
	\end{align}	
	Using a unitary transformation $R=\exp[-i\omega_0( a^\dag a+m^\dag m)t]$, we can change system's operators as $a\rightarrow a\exp(-i\omega_0 t)$, $a^\dag\rightarrow a^\dag\exp(i\omega_0 t)$, $m\rightarrow m\exp(-i\omega_0 t)$, and $m^\dag\rightarrow m^\dag\exp(i\omega_0 t)$. Thus, 
	\begin{align}
	H_{\rm om} &\rightarrow  (\omega_a-\omega_0) a^\dag a+\dfrac{1}{2}\omega_b(p^2+q^2)-g_{0} a^\dagger aq,\notag\\
	H_{\rm kerr}&\rightarrow \Delta_m m^\dag m+K_0 (m^\dag m)^2\notag \\
	H_{\rm int}&\rightarrow H_{\rm int},~~H_D\rightarrow i\Omega_0 (a^\dag-a ),
	\end{align}
	where $\Delta_m=\omega_m-\omega_0$.
	
	With the Heisenberg-Langevin approach, the quantum dynamics of the considered system can be governed by the following quantum Langevin equations:
	\begin{align}\label{HQLE}
	\dot{q}&=\omega_b p,\notag\\
	\dot{p}&=-\omega_b q+g_0 a^\dag a-\gamma_b p+\xi,\\
	\dot{a}&=-[i(\omega_a-\omega_0)+\kappa_a]a+ig_0 a q-ig_m m+\Omega_0+\sqrt{2\kappa_a}a_{\rm in},\notag\\
	\dot{m}&=-(i\Delta_m +\gamma_m)m-ig_m a-i K_0 m^\dag m m+\sqrt{2\gamma_m}m_{\rm in},\notag
	\end{align}
	where $\kappa_a$ and $\gamma_m$ are the decay rates of the cavity and mechanical modes, respectively. $a_{\rm in},~m_{\rm in}$, and~$\xi $ are the input noise operators with zero mean value (i.e., $\langle a_{\rm in}\rangle=\langle m_{\rm in}\rangle=\langle \xi\rangle=0$).
	
	Below we employ the standard linearization method~\cite{Vitali-2007} to derive the linearized Hamiltonian in Eq.~(1) (see the main text). We rewrite each operator as the sum of the mean value (i.e., operator expectation) and the corresponding fluctuation, i.e.,
	\begin{equation}
	q\rightarrow q_s+q,~p\rightarrow p_s+p,~a\rightarrow a_s +a,~m\rightarrow m_s+m.\label{s7}
	\end{equation}
	Substituting Eq.~(\ref{s7}) into Eq.~(\ref{HQLE}), we can obtain the following equations for the mean values of the operators:
	\begin{align}
	\dot{q}_s&=\omega_b p_s,~~\dot{p_s}=-\omega_b q_s+g_0 |a_s|^2-\gamma_b p_s,\notag\\
	\dot{a}_s&=-(i\tilde{\Delta}_a+\kappa_a)a_s-ig_m m_s+\Omega_0,\notag\\
	\dot{m}_s&=-(i\tilde{\Delta}_m +\gamma_m)m_s-ig_m a_s,\label{s8}
	\end{align}
	where $\tilde{\Delta}_a=\omega_a-\omega_0-g_0 q_s$ is the effective cavity frequency detuning induced by the displacement of the mechanical resonator, and $\tilde{\Delta}_m=\Delta_m+\Delta_K$ with $\Delta_K=2K=4K_0 N_m=4K_0 |m_s|^2$ is the effective magnon frequency detuning induced by the Kerr effect. 
	In the long-time limit, $\dot{q}_s=\dot{p_s}=\dot{a}_s=\dot{m}_s=0$. The steady-state condition directly gives
	\begin{align}
	p_s&=0,~q_s=g_0 |a_s|^2/\omega_b,\notag\\
	m_s&=-ig_m a_s/(i\tilde{\Delta}_m +\gamma_m),\\
	a_s&=(\Omega_0-ig_m m_s)/(i\tilde{\Delta}_a+\kappa_a).\notag
	\end{align}
	Also, the equations for the quantum fluctuations, which are obtained by substituting Eq.~(\ref{s7}) into Eq.~(\ref{HQLE}) and neglecting high-order fluctuation terms, are given by
	\begin{align}\label{HLE}
	\dot{q}=&\omega_b p,~~\dot{p}=-\omega_b q+g_b(a+a^\dag)/\sqrt{2}-\gamma_b p+\xi,\notag\\
	\dot{a}=&-(i\tilde{\Delta}_a+\kappa_a) a+ig_b q/\sqrt{2}-ig_m m+\sqrt{2\kappa_a} a_{\rm in},\\
	\dot{m}=&-(i\tilde{\Delta}_m+\gamma_m) m-ig_m a-iKm^\dag+\sqrt{2\gamma_m} m_{\rm in},\notag
	\end{align}
	where $g_b=\sqrt{2}g_0 a_s$ is the enhanced optomechanical coupling by the strong driving field. Obviously, Eq.~(\ref{HLE}) is the same as Eq.~(\ref{q2}) in the main text. The corresponding linearized Hamiltonian of the hybrid system without dissipation can be written as
	\begin{align}\label{Q1}
	\mathcal{H}=&\tilde{\Delta}_a a^\dag a+\tilde{\Delta}_m m^\dag m+\frac{1}{2}\omega_b (q^2+p^2)-\frac{1}{\sqrt{2}}{g_b}(a+a^\dag)q\notag\\
	&+g_m (a^\dag m+a m^\dag)+\frac{1}{2}K(m^{\dag 2}+m^2),
	\end{align}
	which is just Eq.~(\ref{q1}) in the main text. 
	
	By further defining quadratures, 
	\begin{align}
	x_a(t)=&\frac{a+a^\dag}{\sqrt{2}},~y_a(t)=\frac{a-a^\dag}{i\sqrt{2}},\notag\\
	x_m(t)=&\frac{m+m^\dag}{\sqrt{2}},~y_m(t)=\frac{m-m^\dag}{i\sqrt{2}},
	\end{align}
	Eq.~(\ref{HLE}) can be rewritten as
	\begin{align}\label{s11}
	\dot{q}=&\omega_b p,~~\dot{p}=-\omega_b q+g_bx_a-\gamma_b p+\xi,\notag\\
	\dot{x}_a=&-\kappa_a x_a+\tilde{\Delta}_a+g_m y_m+\sqrt{2\kappa_a}x_{\rm in}^a(t),\notag\\
	\dot{y}_a=&-\tilde{\Delta}_a x_a-\kappa_a y_a-g_m x_m+\sqrt{2\kappa_a}y_{\rm in}^a(t),\\
	\dot{x}_m=&-\gamma_m x_a+\tilde{\Delta}_m^- y_m+g_m y_a+\sqrt{2\gamma_m}x_{\rm in}^m(t),\notag\\
	\dot{y}_m=&-\tilde{\Delta}_m^+ x_m-\gamma_m y_m-g_m x_a+\sqrt{2\gamma_m}y_{\rm in}^m(t),\notag
	\end{align}
	where $x_{\rm in}^a(t)=(a_{\rm in}+a_{\rm in}^\dag)/\sqrt{2}$, $y_{\rm in}^a(t)=(a_{\rm in}-a_{\rm in}^\dag)/i\sqrt{2}$, $x_{\rm in}^m(t)=(m_{\rm in}+m_{\rm in}^\dag)/\sqrt{2}$, and $y_{\rm in}^m(t)=(m_{\rm in}-m_{\rm in}^\dag)/i\sqrt{2}$. Here $\tilde{\Delta}_m^\pm=\tilde{\Delta}_m\pm \Delta_K/2$. In a compact form, Eq.~(\ref{s11}) can be given by
	\begin{align}
	\dot{u}(t)=Au(t)+f(t),
	\end{align}
	where $u(t)=[x_a (t), y_a (t), x_m (t),y_m (t),  q(t),  p(t)]^T$ and $f(t)=[\sqrt{2\kappa_a} x_{\rm in}^{a}(t), \sqrt{2\kappa_a} y_{\rm in}^{a}(t), \sqrt{2\gamma_m} x_{\rm in}^{m}(t),
	\sqrt{2\gamma_m} y_{\rm in}^{m}(t)$, $0, \xi(t)]^T$ are the vectors of the system and the input noise operators, respectively, and the drift (coefficient) matrix $A$ is given by Eq.~(\ref{q3}) in the main text.

\end{document}